\documentclass[aps,prc,reprint,groupedaddress]{revtex4-2}
\usepackage{graphicx}  
\usepackage{dcolumn}
\usepackage{subfigure}
\usepackage{hyperref}
\begin{document}
	
\title{Normalized factorial moments of spatial distributions of particles in high multiplicity events: A Toy model study}
\author{Sheetal Sharma, Salman K. Malik, Zarina Banoo} 
\author{Ramni Gupta}\thanks{}
\email[]{ramni.gupta16@gmail.com}
\affiliation{Department of Physics, University of Jammu, Jammu, India}
\date{\today}

\begin{abstract}

 In ultra-relativistic heavy-ion collisions a strongly interacting complex system of quarks and gluons is formed. The nature of the system so created and the mechanism of multi-particle production in these collisions may be revealed by studying the normalized factorial moments ($F_{{\rm{q}}}$) as function of various parameters. The resilience of $F_{{\rm{q}}}$ moments studied using Toy model events shows that these are sensitive to the presence of dynamical fluctuations in the system and are robust against the uniform efficiencies in the data measurements. Results of this study serve as a suitable reference baseline for the experimental and simulation studies.
\end{abstract}
\date{\today}
\maketitle
\section{Introduction}
\label{intro}
The Large Hadron Collider (LHC) at CERN, Switzerland \cite{ALICE:2010suc}, and the Relativistic Heavy-Ion Collider (RHIC) at BNL, USA~\cite{Gyulassy:2004zy}, serve the purpose of studying the deconfined state of strongly interacting particles, the Quark-Gluon Plasma (QGP) and unraveling its properties by colliding nuclei at ultra-relativistic energies \cite{VanHove:1983rk, Harris:1996zx}. This state of matter can be obtained by colliding heavy nuclei at high energies such that the energy density and temperature melts hadrons \cite{Busza:2018rrf}. As the medium created cools down, a transition occurs from the QGP state to the hadronic state. One of the aims of these collider experiments is to probe the phase diagram of this strongly interacting matter and its properties \cite{Hands:2001ve, Akiba:2015jwa, Shuryak:1980tp}.

In the phase diagram of a system, at a critical point, the correlation length of the system becomes infinite and scale invariant~\cite{Stephanov:2004wx}. A study of critical behavior is vital to have a deeper understanding of the properties of any system and knowledge about the nature of the phase transition between different phases. At this point, the system exhibits large fluctuations in various observables~\cite{Bluhm:2020mpc, NA61SHINE:2015uhh}, and a study of these provide a powerful means to understand the myriad characteristics of the system.

Some of the features of the matter may only get revealed at very high energy density~\cite{Kittel:b}.  The charged particle multiplicity in heavy-ion collisions is a function of the energy and temperature. Fluctuations in the multiplicity distributions are one of the main observables of these collision experiments. In~\cite{Hwa:2011bu},  it is suggested that the normalized factorial moments (NFMs) of the charged particle multiplicity distributions recorded using the colliders, as at LHC, be analyzed for the study of bin-to-bin and event-to-event multiplicity fluctuations.

A power-law behavior of the normalized factorial moments of the particle density fluctuations in spatial or momentum space with an increasing number of bins is termed as intermittency \cite{Bialas:1985jb, Bialas:1990gt, Wolf}. This analysis technique has been performed for various systems at low energies in search of QGP and to understand the quark-hadron phase-transition~\cite{Kittel:b}. However, investigations of heavy-ion collision data using this tool have led to no definitive conclusions on the critical point, order of  phase transition, or nature of multiplicity fluctuations~\cite{Wolf} because of low bin multiplicities. The availability of data with high charged particle density per bin has generated a renewed interest, such as at the STAR experiment  at RHIC~\cite{STAR:2023jpm}, ALICE at CERN~\cite{qmposter}, to use this methodology to comprehend the multiparticle production processes. 

A detailed study of the normalized factorial moments of the spatial distributions in high multiplicity events generated using the Toy model is reported here in extension to~ \cite{Sharma:2021cyp}. In section~\ref{evt_gen} an introduction of the Toy model event generation followed by details on the methodology of analysis is given. Observations and results of the analyses are discussed in section~\ref{result} and a summary of the work is given in section~\ref{sum}.

\section{Event generation and analysis methodology}
\label{evt_gen}
Processes leading to multiparticle production in heavy-ion collisions are still not known.  A number of methodologies  and event generators developed using theories and models are studied to investigate particle production mechanisms to find answers to these mysteries.  In high-energy physics research event generators are extensively employed by experimentalists to simulate experimental conditions and to understand outcomes from the experiments based on known physics implemented in the models.   Here, a baseline behavior of the normalized factorial moments, in the contours of the methodology proposed in~\cite{Hwa:2011bu} is investigated using events generated with a Toy model, for its suitability to analyze heavy-ion collision data. 

\begin{figure}[htp]
	\centering
	\includegraphics[width=8.0cm,height=6.5cm]{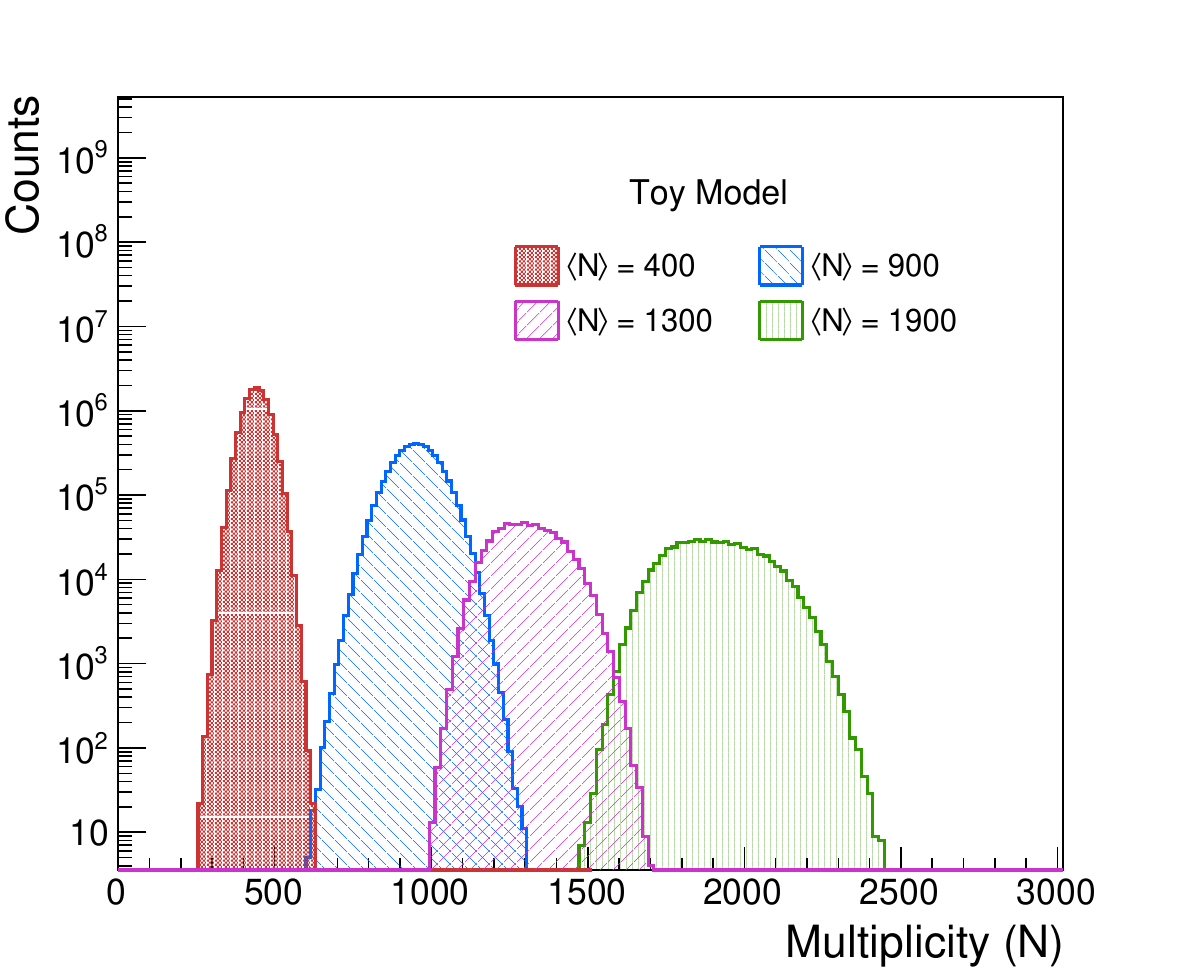}
	\caption{Multiplicity distributions for the four sets of event samples generated using Toy model.}
	\label{etaphi}
\end{figure}

Intermittency analysis, first proposed in 1986 \cite{Bialas:1985jb} to investigate fluctuations in the pseudorapidity distributions of some cosmic ray events, is suggested \cite{Hwa:1992uq, Hwa:2011bu, Hwa:2014jea} to be performed for high multiplicity data recorded by the detectors in the recent colliders, such as at the LHC, to understand the multiparticle production and quark-hadron phase transition. The ALICE experimental setup at LHC has a central barrel detector that includes TPC, ITS, and TOF detectors. These detectors  together measure the charged particles within a common angular phase space of pseudorapidity  ($|\eta|\!\leq$ 0.8) and full azimuthal angle ($0 \!\leq \!\varphi \! \leq \!2\pi$) \cite{ALICE:2008ngc, ALICE:2016tlx}.  Toy model event samples  are generated with charged particle multiplicity distributions similar to the one recorded for midrapidity region by ALICE detector for low $p_{\rm{T}}$ tracks ($<\!2.0$~GeV/c). Four event samples with multiplicity distributions, as shown in Fig.~\ref{etaphi}, having mean multiplicity 400, 900, 1300, and 1900 are simulated using global random generator  object (gRandom) provided by the ROOT~\cite{Brun:1997pa} analysis framework, using system clock as seed. Uniformly distributed particles are generated in the spatial phase space $(\eta,\varphi)$, having $|\eta|\,\leq$ 0.8 and  $0 \!\leq \!\varphi \! \leq \!2\pi$ such that there are no bin-to-bin (spatial) fluctuations.

\begin{figure}
  \includegraphics[width=0.45\textwidth]{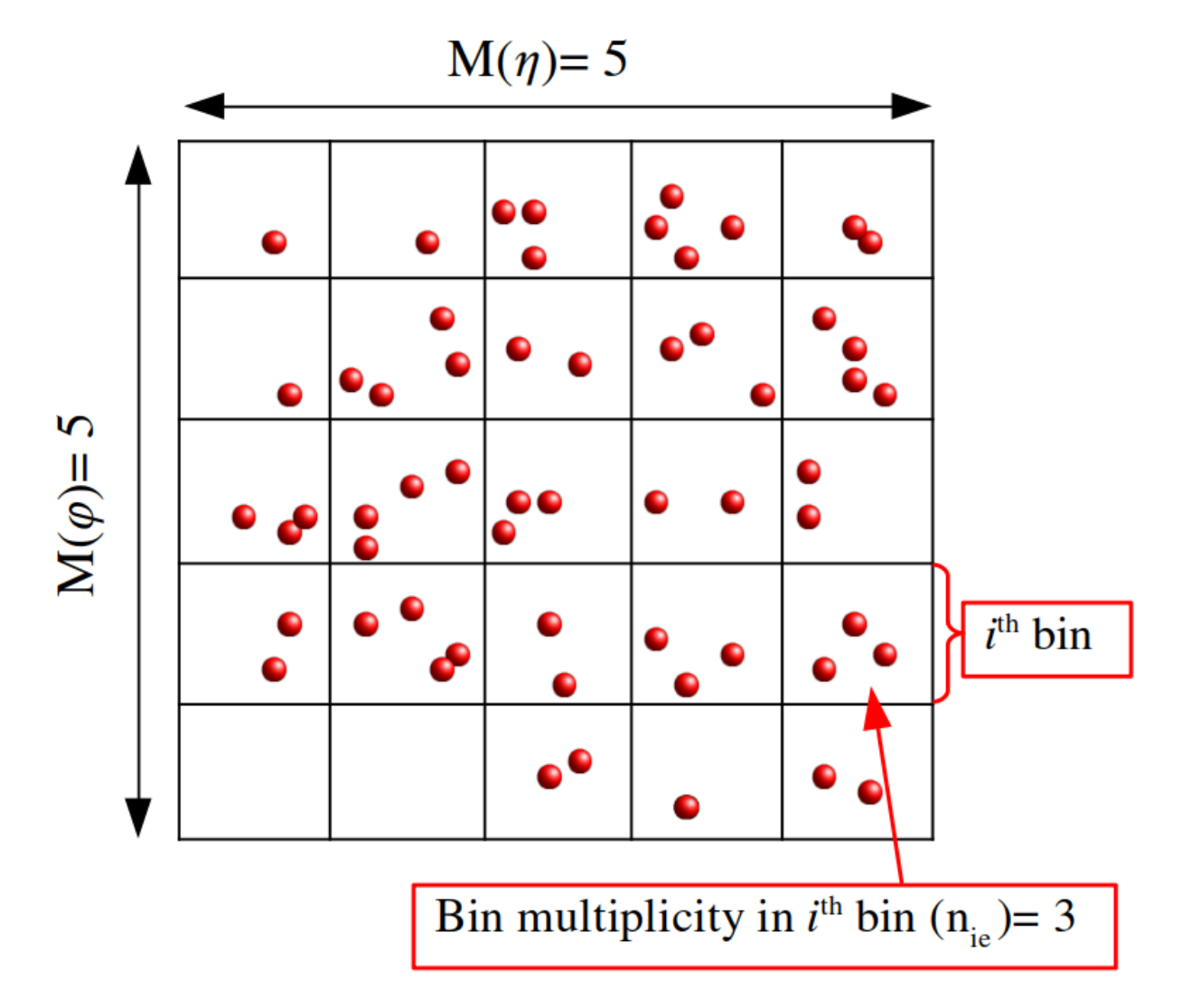}
\caption{Graphic illustration showing two dimensional ($\eta,\varphi$) phase space partitioned in M$\,\times\,$M bins and mapping of tracks of an event onto this partitioned phase space region (grid) is also shown.}
 \label{binning}
\end{figure}

Lattice QCD predicts large density fluctuations in a system at  a critical point while undergoing phase transition. It is argued in \cite{Hwa:2011bu} that the behavior of NFM as a function of the number of bins is a good measure of fluctuations in the spatial configurations of produced particles in the heavy-ion collisions. This methodology, similar to the intermittency analysis as proposed earlier
in \cite{Bialas:1985jb, Bialas:1988wc, Bialas:1990gt, Bialas:1990vh}, has been tested for some of the models at LHC energies \cite{Gupta:2019zox, Sharma:2021cyp}. In present work, a two-dimensional (2D) analysis is performed for the Toy model events wherein for an event the kinematic phase space is partitioned into a square lattice with $M_{{\rm\eta}}$ and  $M_{\rm{\varphi}}$ number of bins along $\eta$ and $\varphi$, respectively, as is depicted in Fig.\ref{binning}. With $M_{{\rm{\eta}}}\! = \!M_{{\rm\varphi}}$ there are $M^{2}$  number of bins, such that $q^{\rm{th}}$ order normalized factorial moments (NFM), for a total of $N$ events, is defined as
\begin{equation}
F_{\rm{q}}(M)= \frac{\displaystyle \frac{1}{N} \displaystyle\sum_{e=1}^{N}\frac{1}{M^{2}}\sum_{i=1}^{M^{2}}f_{q}^{e}(n_{\rm{ie}})}{\left (\displaystyle\frac{1}{N} \displaystyle \sum_{e=1}^{N}\frac{1}{M^{2}}\sum_{i=1}^{M^{2}}f_{1}^{e}(n_{\rm{ie}}) \right )^q} 
\label{def}
\end{equation}
where
\begin{equation}
     f_{\rm{q}}^{e}(n_{\rm{ie}})= \Pi_{j=0}^{q-1}(n_{\rm{ie}}-j),
    \label{fm}
\end{equation}
and $n_{\rm{ie}}$ is the bin multiplicity (number of particles)  in the $i^{\rm{th}}$ bin of the $e^{\rm{th}}$ event. $q$ takes positive integer values $\geq 2$. $M$ is varied from 4 to 100, and $q$  from 2 to 5 in integral steps. By definition, these normalized factorial moments ($F_{\rm{q}}$(M)) filter out statistical fluctuations. If fluctuations in the spatial distributions of the particles are Poissonian, then $F_{q}$(M)\,=\,1  \cite{Bialas:1985jb}. A power-law behavior of $F_{q}$ for $q\geq 2$ as function of $M^{\rm{D}}$ (D signifies the phase space dimensions, in the present analysis  D=2)
\begin{equation}
F_{\rm{q}}(M) \propto {(M^{2})}^{\phi_{\rm{q}}}
\label{eq:mscal1}
\end{equation}
then defines intermittency (also termed as M-scaling), where $\phi_{\rm{q}} > 0$ is known as intermittency index or scaling index. Dependence of $F_{\rm{q}}(M)$, for q $>$ 2, on the second order factorial moment ($F_{\rm{2}}(\rm{M})$) is proposed in~ \cite{Hwa:1992uq, Ochs:1988ky}, where the  intermittency is studied within the framework of Ginzburg-Landau formalism, such that
\begin{equation}
F_{\rm{q}}(M) \propto (F_{\rm{2}}(M))^{\beta_{\rm{q}}}, 
\label{fqx}
\end{equation}
\begin{figure}[hbt!]
\centering
\includegraphics[width=0.23\textwidth]{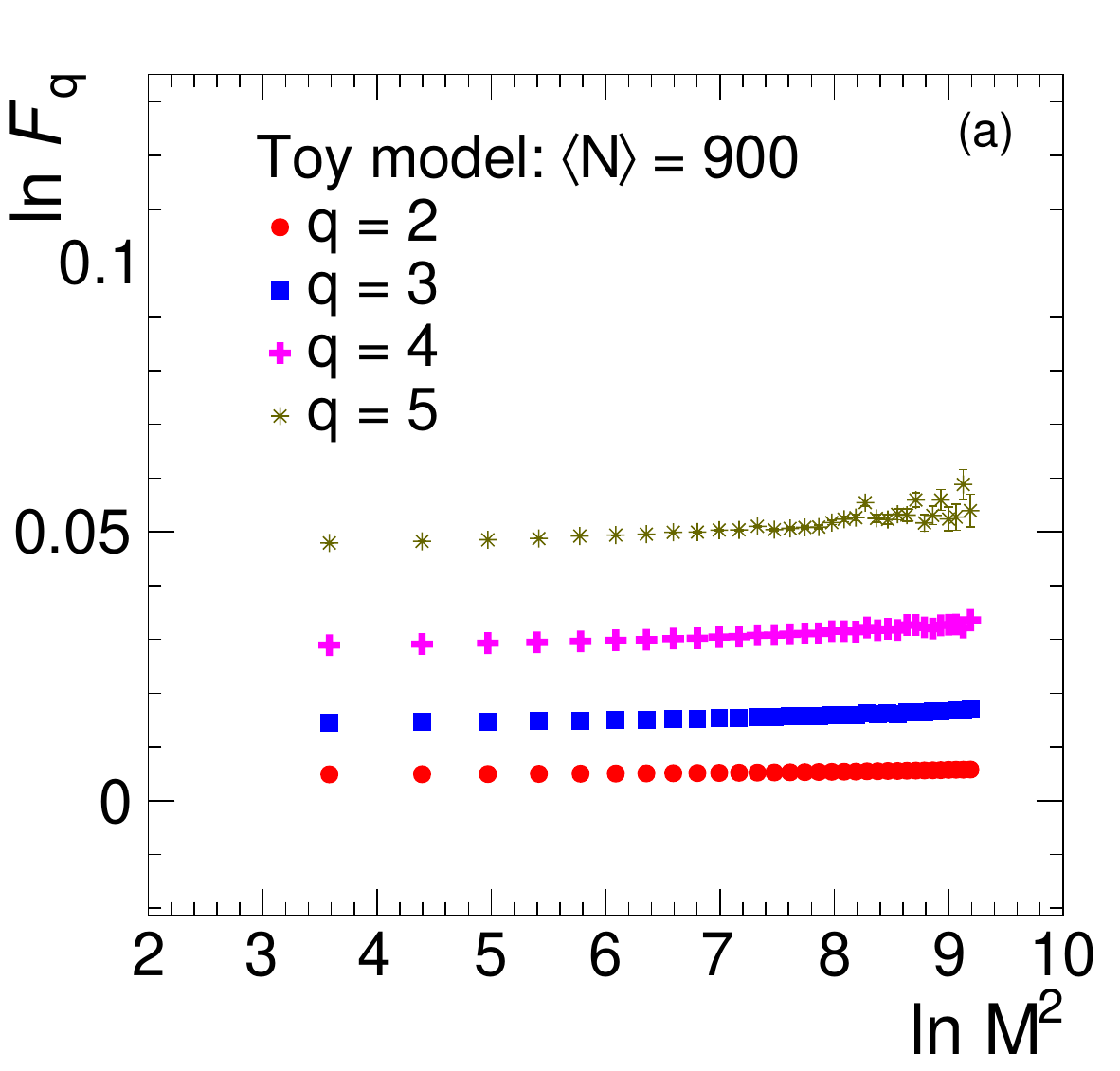} 
\includegraphics[width=0.23\textwidth]{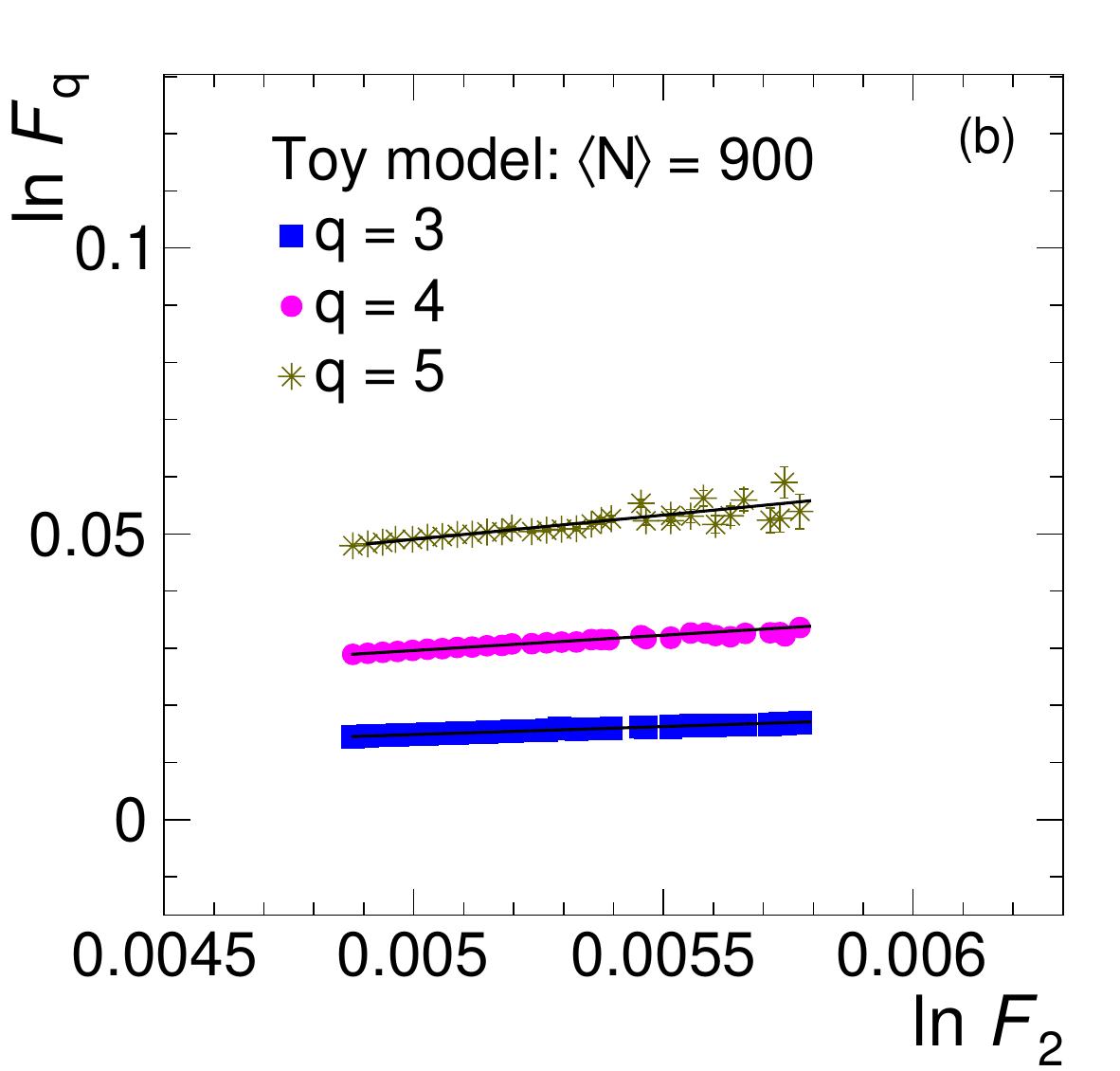}
\caption{(Left) log-log plot of $F_{\rm{q}} (M)$ vs $M^{2}$ (M-scaling) for q = 2, 3, 4 and 5; (Right) log-log plot of $F_{\rm{q}} (M)$ vs $F_{\rm{2}}$ (F--scaling) for q = 3, 4 and 5 for the event sample with  mean multiplicity $\langle N \rangle = 900$. Similar observations are made for the other event samples studied.}
\label{fig:result1}
\end{figure}
and $\beta_{q}$ is the exponent which depends on the bin resolution and number of bins $M$. This scaling is termed as F-scaling. $\beta_{q}$ and $\phi_{q}$ depend on different critical parameters of the system. Thus, F-scaling is independent of the M-scaling behavior.  Relating  to order of the moment ($q$), a scaling exponent ($\nu$) can be obtained from the relation
\begin{equation}
\beta_{\rm{q}} = (q-1)^{\nu}.
\label{betax}
\end{equation}
The scaling exponent is independent of the specific values of critical parameters and serves as a tool to examine the occurrence of QCD phase transition \cite{Hwa:1992uq, Hwa:2011bu, Hwa:1992ii}. Theoretical predictions  from Ginzburg-Landau (GL) theory \cite{Hwa:1992uq} for the value of $\nu$  is 1.304, and 1.0 based on calculations derived from a two-dimensional Ising model \cite{Hwa:1992ii}. The scaling behavior and sensitivity of NFM for the Toy model events is investigated. In the backdrop of various detector effects that may distort the original signal the efficiency correction method, required to be implemented  for the extraction of true behavior of NFM  is also studied. 
\begin{figure}[hbt!]
	\centering
	\includegraphics[width=0.23\textwidth]{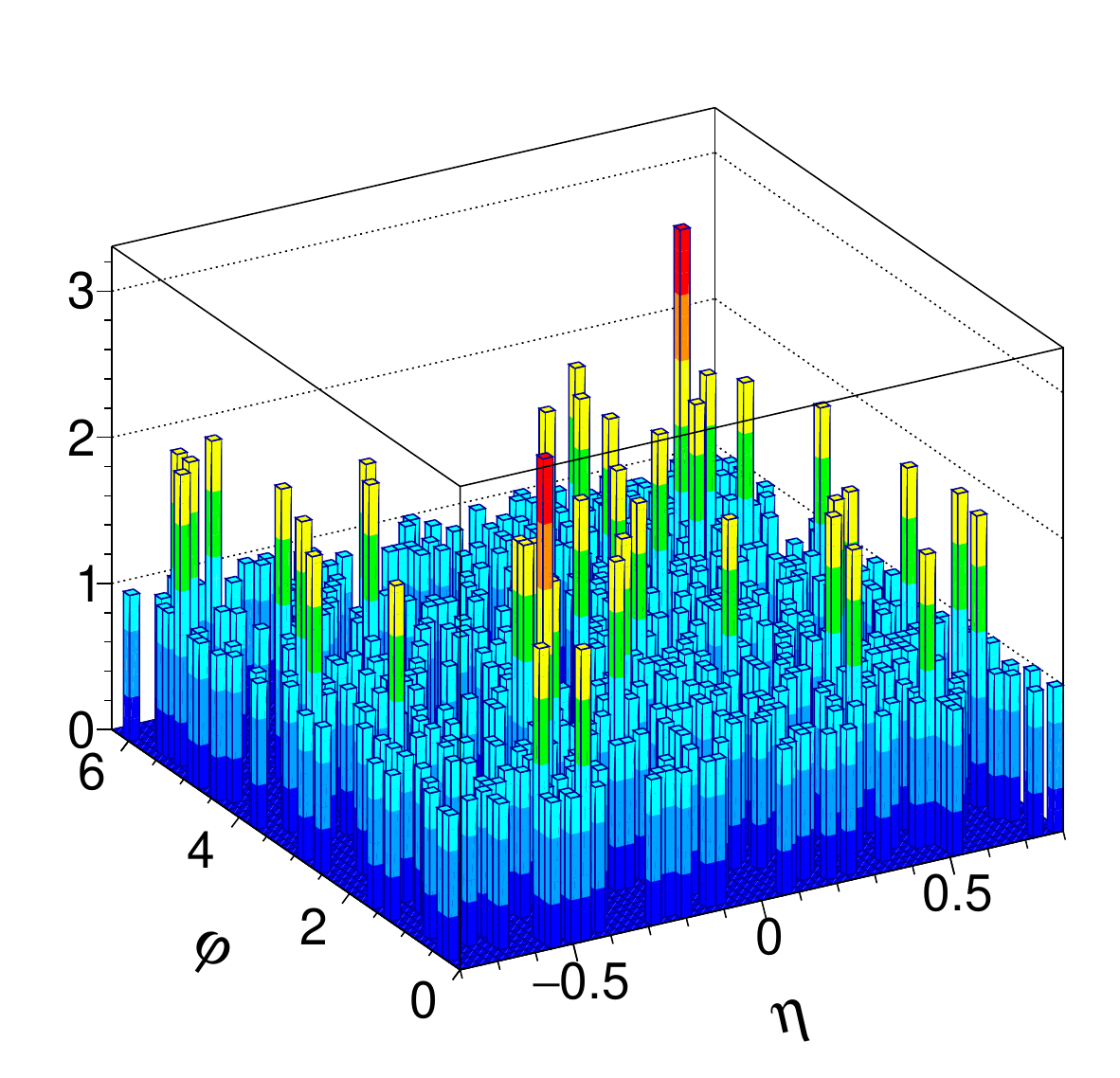}
	\includegraphics[width=0.23\textwidth]{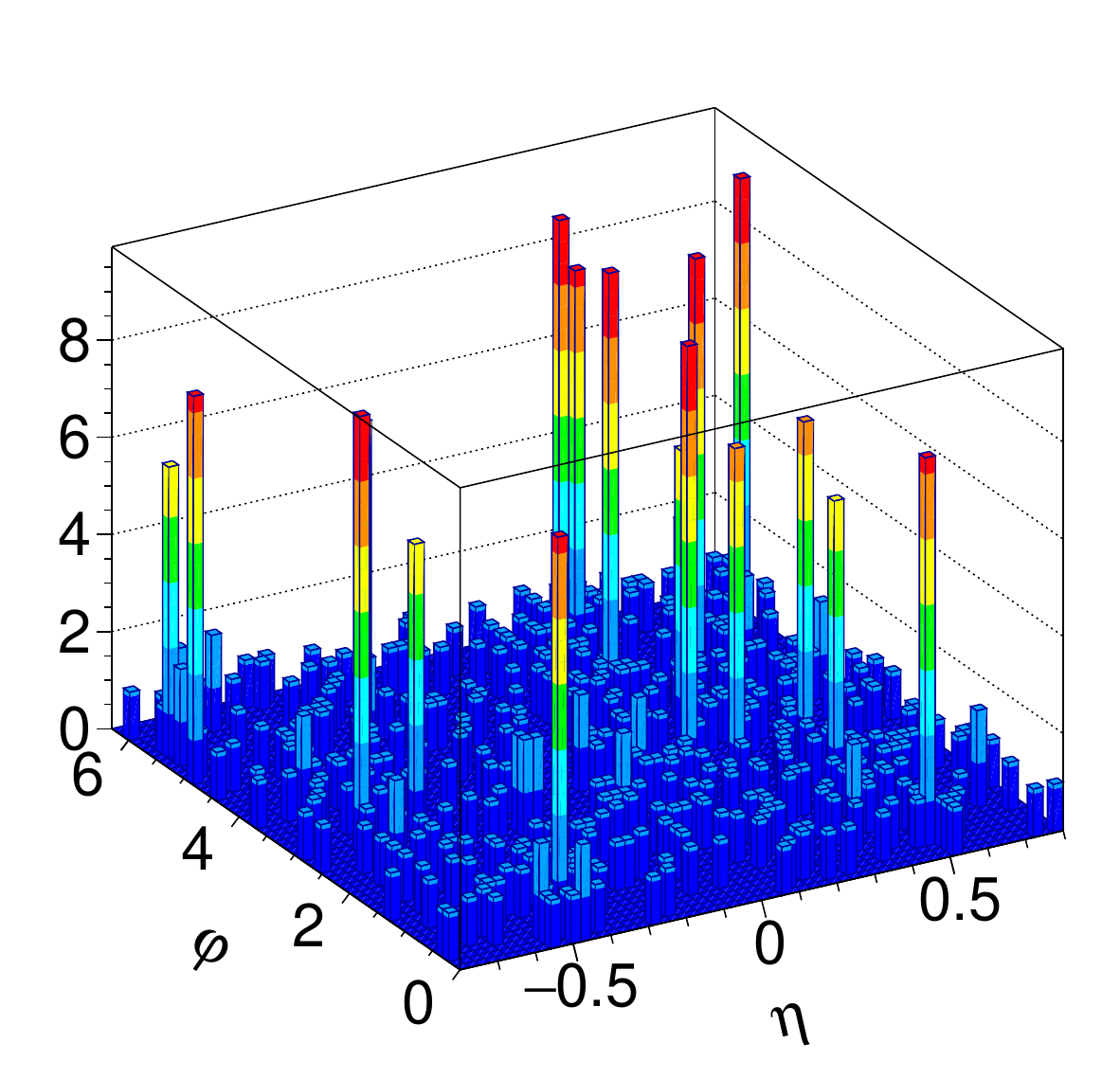}
	\caption{(Left plot) Spatial distribution of particles in the ($\eta,\varphi$) phase space of an event. Panel on right shows the spatial distribution of particles of the same event with 5\% tracks added in selected bins while removing equal number of tracks from rest of the region (modified Toy model events).}
	\label{int1}
\end{figure}
\section{Observations and Discussion} 
\label{result}
\subsection{Scaling of normalized factorial moments (NFM)}
An event-by-event analysis is performed where  an event is mapped onto a two-dimensional phase space partitioned into $M^{2}$ bins. Event factorial moments are  determined  for q = 2, 3, 4, and 5 using Eq.~\ref{fm}. Normalized factorial moments  ($F_{q=2,3,4,5}(M)$) are then calculated for the whole event sample, as defined in Eq.~\ref{def}. Fig.~\ref{fig:result1}(a) shows the dependence of $F_{q}$ on $M^{2}$ in a log-log plot for an event sample with average multiplicity of 900. Statistical uncertainties on the data points are estimated using the  sub-sampling method. 

For the second order NFM  $F_{\rm{q}}(M)$ is observed to be  $>\!1$. As the value of $q$ increases, the deviation of $F_{q}$ from 1 increases.  However, $F_{q}$ shows no dependence on $M$ (no power-law as in Eq.~\ref{eq:mscal1}), i.e., as $M$ increases there is negligible change  in the $F_{\rm{q}}$ (for all $q{\rm{'s}}$) and thus no intermittency in the generated events.  No significant rise in $\ln{F_{\rm{q}}}$ with $\ln{M^{2}}$ seen for the Toy model indicates the  absence of self-similar modeling in particle generation.  F-scaling (Eq.~\ref{fqx}),  a $\ln\! F_{q}(M)$ versus $\ln\!F_{2}(M)$ plot,  for the same event sample is given in Fig.~\ref{fig:result1}(b). For $q>2$, $F_{q}$ shows linear dependence on $F_{\rm{2}}$ and line fit to these graphs give slopes, $\beta_{q}$. A dimensionless scaling exponent (Eq.~\ref{betax}), which quantitatively characterizes the geometric spatial configurations of the Toy model events, obtained using these slope values, is discussed in section~\ref{scalexpo}.
\subsection{Sensitivity of NFM to gauge fluctuations}
The independence of $F_{\rm{q}}$(M)  on the number of bins ($M^{2}$) for the Toy model events, as observed above, can be because of two possible reasons. One can be that the observable $F_{\rm{q}}(M)$ are not sensitive to multiplicity fluctuations  from bin-to-bin and the second possible reason which can lead to this observation can be that there are no bin-to-bin fluctuations in the events. The Toy model events have been simulated for uniform spatial distributions with no bin-to-bin fluctuations.  Thus, the behavior of NFMs is required to be investigated in the presence of fluctuations. For investigating the sensitivity of the observable to gauge the density fluctuations in the spatial configurations, fluctuations are added with hand (termed hereafter \textit{artificial fluctuations}) in each event, and the scaling behavior of normalized factorial moments is studied. 

The event sample that results after including artificial fluctuations in each Toy model event  is then termed as {\textit{ modified Toy model event}}. For getting such an event with artificial fluctuations, a number of tracks equal to 5\% of the multiplicity of an event, are added in some region of the phase space of the event. At the same time, an equal number of tracks are removed from the rest of the region to keep the multiplicity distribution of the modified Toy model events same as that of the original Toy model event. This makes some of the bins in the phase space, more populated than the rest of the bins, thereby including multiplicity spikes in the spatial configurations which replicates a system  having  critical fluctuations. Fig.~\ref{int1}, depicts ($\eta$, $\varphi$) lego plot of one  such  Toy model event (left panel) in which  the phase-space is partitioned into 40 bins and the same event after it is modified with artificial fluctuations is also shown (right panel).
\begin{figure}[hbt!]
  \centering
  \includegraphics[width=0.23\textwidth]{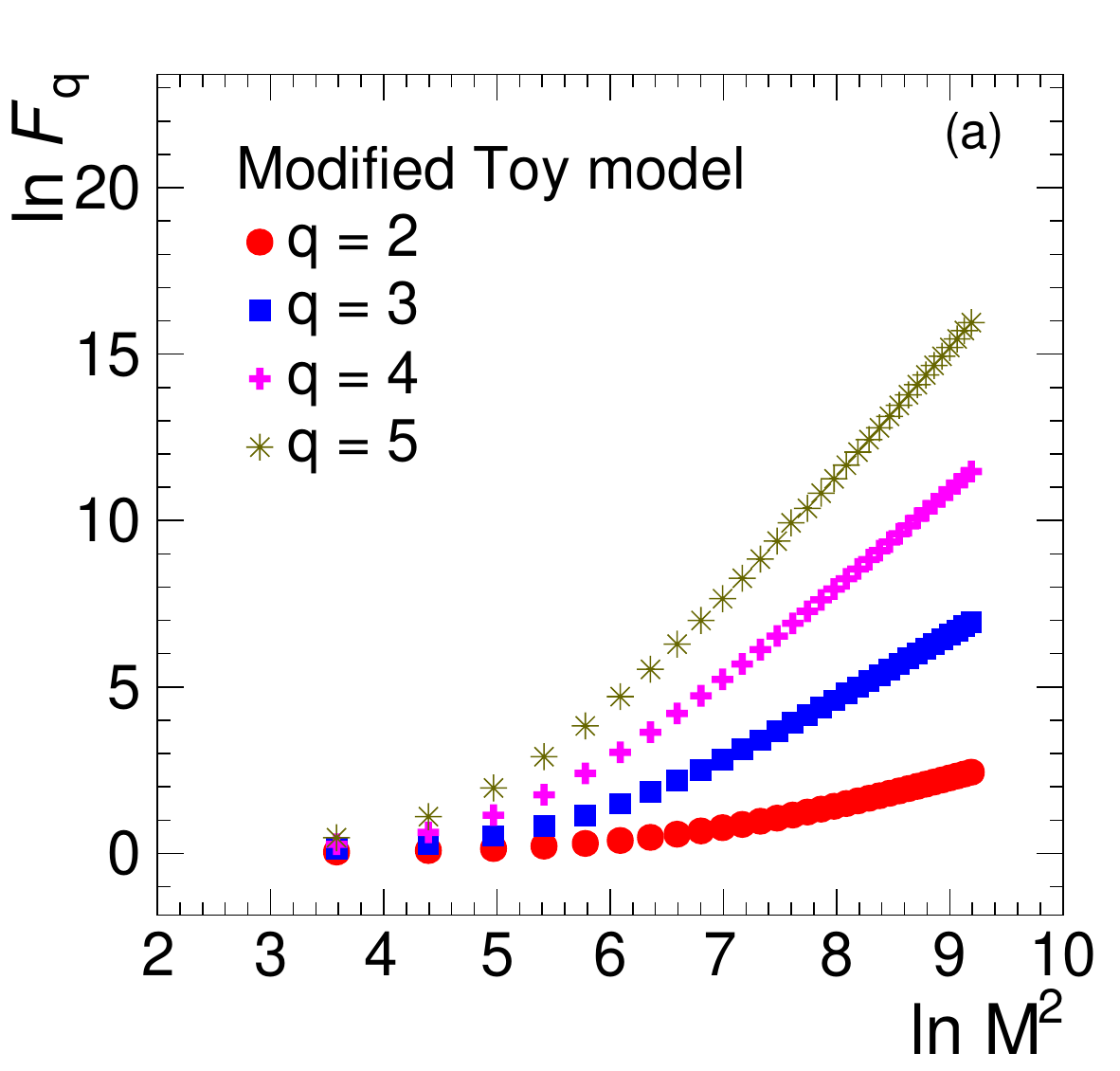}
  \includegraphics[width=0.23\textwidth]{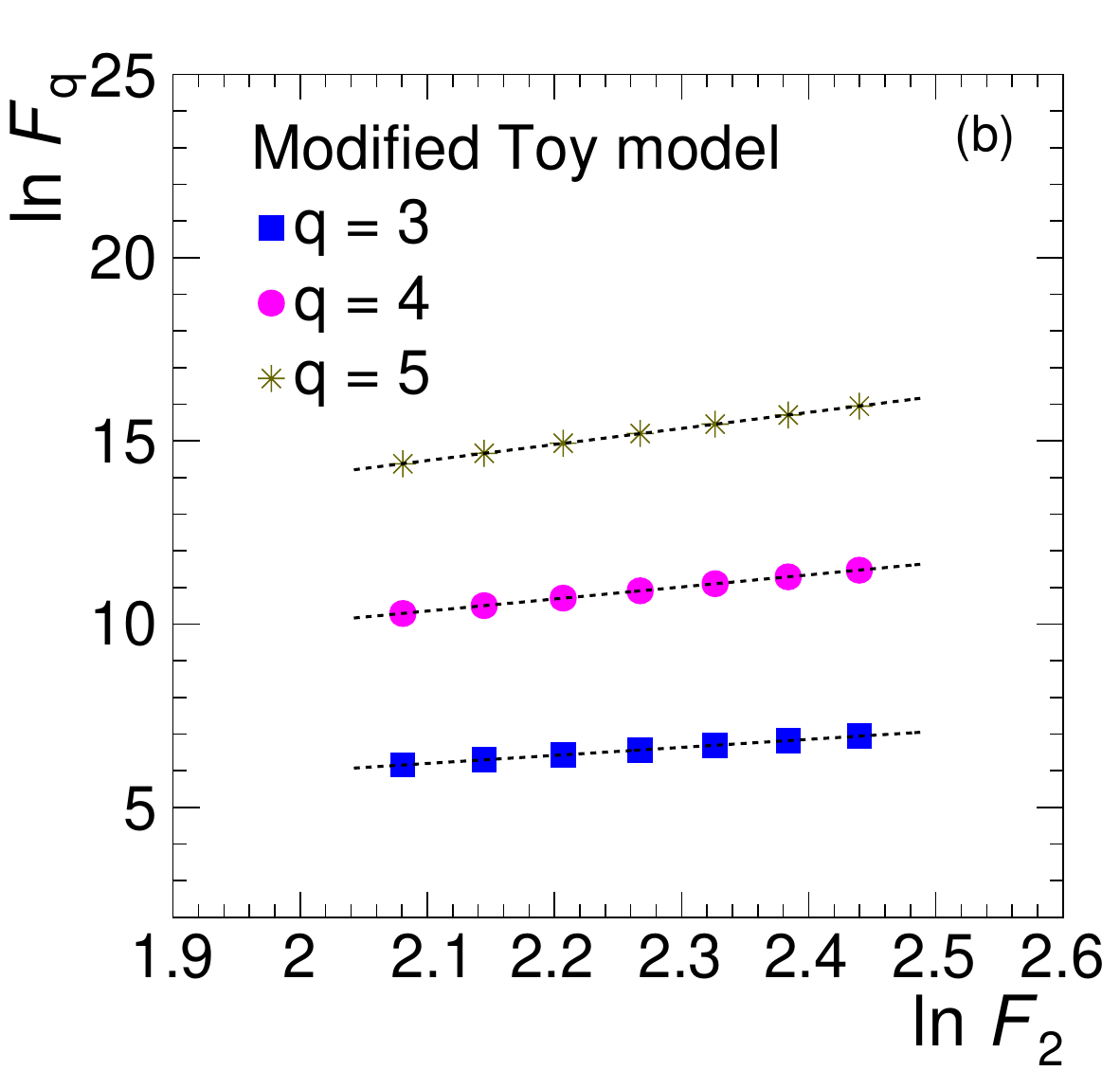}
   \caption{ log-log plot of (a) $F_{\rm{q}}$(M) vs $ M^{2}$ (M-scaling) for q = 2, 3, 4 and 5 and (b) $F_{\rm{q}}$(M) vs $F_{\rm{2}}$(M) (F--scaling) for q = 3, 4 and 5}
  \label{int3}
\end{figure}

Analysis is performed on the modified Toy model events, and normalized factorial moments are determined for q = 2, 3, 4 and 5 with M varying from 4 to 100. The M-scaling behavior is studied, where in  Fig.\ref{int3}(a) log-log plot between $F_{\rm{q}}$(M) and $M^{2}$ is given.  A power-law dependence of $F_{\rm{q}}$(M) on $M^{2}$ is observed. For  high $M$ values  $F_{\rm{q}}$ $>>$ 1 and this deviation increase as M increases. These results show that NFM are sensitive to bin-to-bin multiplicity fluctuations and thus are suitable measure to gauge fluctuations. 
\subsection{Scaling exponent ($\nu$) and its dependence on multiplicity}
\label{scalexpo}
\begin{figure}
    \centering
    \includegraphics[width=0.37\textwidth]{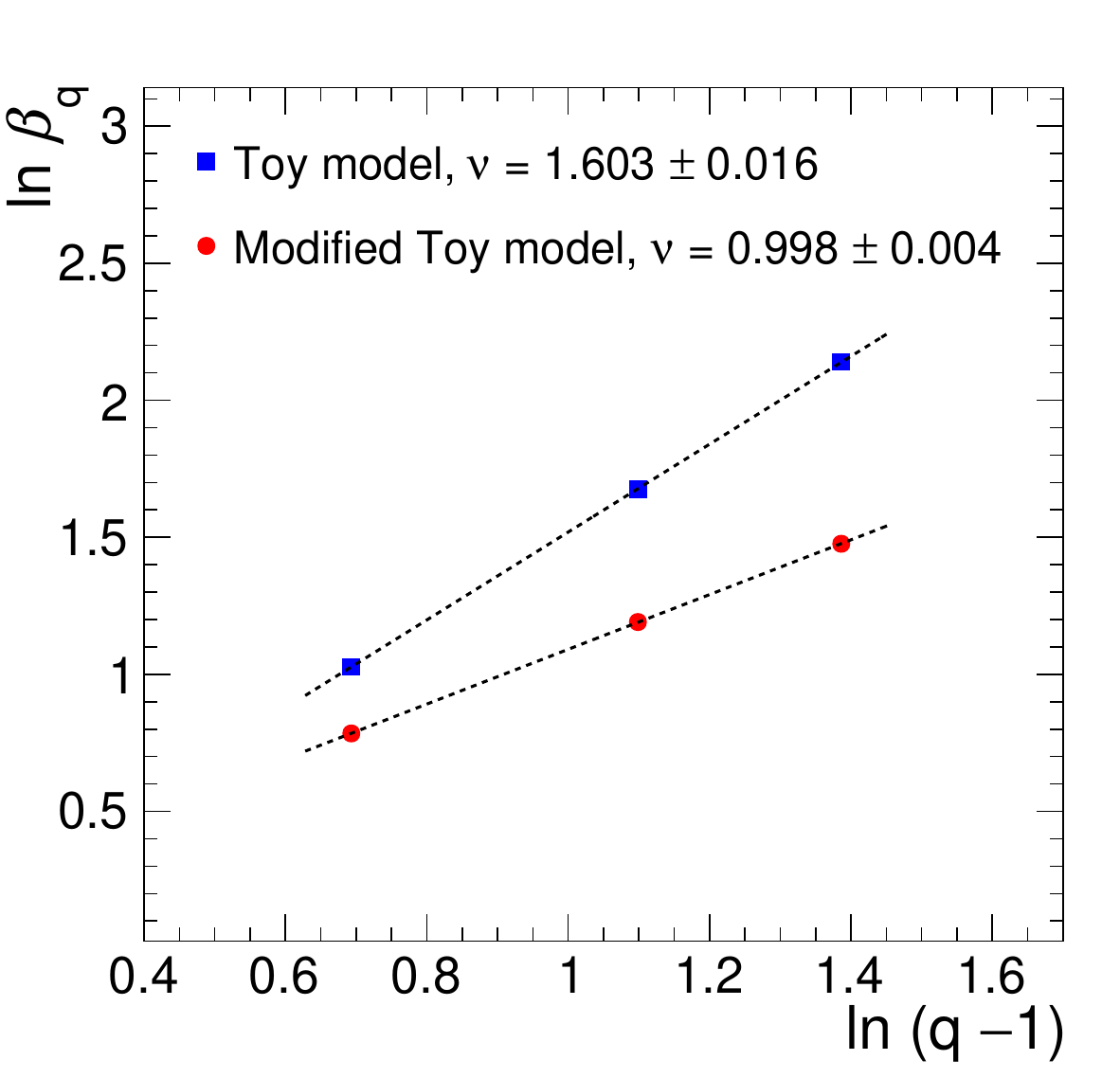}
    \caption{ Scaling exponent ($\nu$) obtained from line fit to $ln \! \beta_{\rm{q}}$ vs $ln (q-1)$ plot for the Toy model events and modified Toy model events.}
    \label{nu}
\end{figure}

Scaling exponent ($\nu$) obtained from a line fit to the  $\ln \beta_{q}$ vs $\ln (q-1)$ plot (Eq.~\ref{betax}) provides valuable insight into the underlying physics of the heavy-ion collisions~\cite{Hwa:1992uq}. Scaling exponent for the Toy model events, with no bin-to-bin fluctuations, is  $1.603\pm0.016$. In case of modified Toy model events the value of $\nu$ is $0.998\pm0.004$ as is shown in Fig.~\ref{nu}.  Toy model events having no preferred bins for track generation, a little addition of tracks in a small region of phase space, alters bin multiplicity and gives large fluctuations in spatial patterns with  $\nu \approx 1.00$, a value which is observed in case of two-dimensional Ising model with large fluctuations ~\cite{Kittel:b}. 

\begin{figure}
\centering
\includegraphics[width=0.37\textwidth]{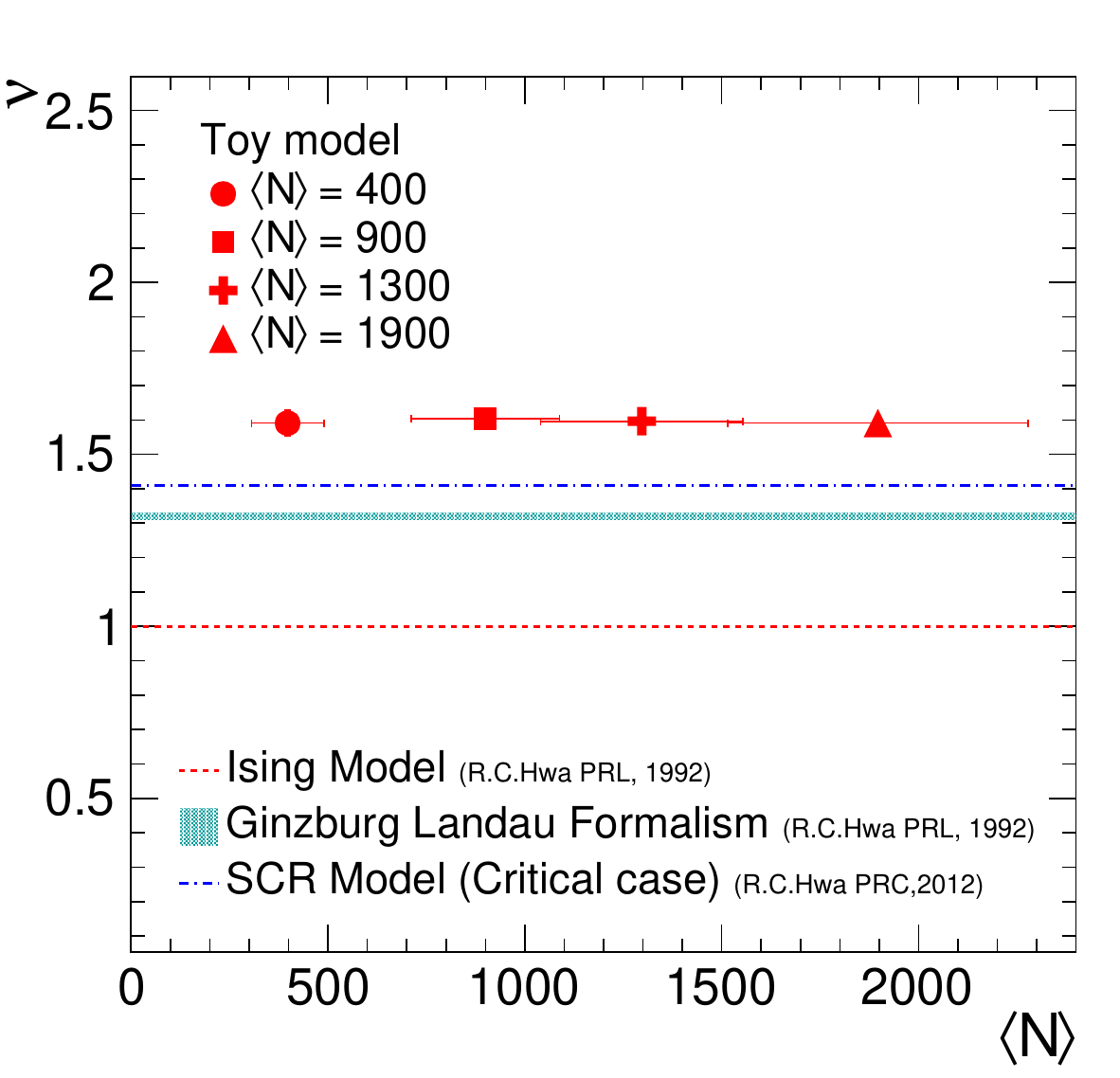}
\caption{Scaling exponent as function of average multiplicitiesof the four sets of Toy model events. Scaling exponent values predicted from the other models are also shown.}
\label{nuvsmult}
\end{figure}

Multiplicity increases with the increase in collision energy. It would be interesting to know whether there is any multiplicity and hence energy dependence of the scaling exponent  ($\nu$).  To investigate this, events with multiplicity distributions, having value of the mean varying from 300 to 1900 (Fig.~\ref{etaphi}),  are generated  and analyzed to study scaling behavior.  Scaling exponent ($\nu$)  is observed to be  independent of the average multiplicity as is shown in Fig.~\ref{nuvsmult}, where the values predicted from models and theory are also shown.  The Toy model events which do not have any bin-to-bin fluctuations the scaling exponent is different from the one predicted for the systems with critical fluctuations and is also independent of the number of particles produced (multiplicity). This signifies the importance of the scaling exponent that is a pure number which quantifies the characteristic inherent fluctuations in the system.
\subsection{Resilience of NFM to efficiency corrections}
In heavy-ion collisions, particles produced are recorded by detectors which may have some inefficiencies in detecting all the particles of interest. These limitations stem from non-optimal detector resolution and inefficiencies in tracking routines. The measurements and calculations of the observable are thus affected by these inefficiencies and need to be corrected before making any reasonable conclusions. Monte Carlo event generators may be used to calculate overall efficiencies.  For this, model generated events are passed through the detector geometry and then recorded and track reconstruction routines are applied  as in the case of experimental setup. Detector efficiency is the ratio of the number of tracks detected by the detector geometry to the number of generated tracks in the acceptance region. Constrained by different conditions, detector efficiencies in the acceptance region can be binomial or non-binomial. The inefficiencies may alter true values of the normalized factorial moments and hence also their behavior with the bin resolution or the number of bins ($M^{2}$). The two-dimensional efficiency maps are determined for each $M$ value. If $\epsilon_{i}$ denotes the efficiency of the $i^{\rm{th}}$ bin of the partitioned phase space, then the normalized factorial moments, as defined in Eq.(\ref{def}), are corrected for the efficiencies as
\begin{equation}
F_{\rm{q}}^{corr.}(M) = \frac
                    { \displaystyle \frac {1}{N} \displaystyle\sum_{e=1}^{N}\frac{1}{M^{2}}\sum_{i=1}^{M^{2}} \frac {f_{q}^{e}(n_{\rm{ie}})} {{\epsilon_{i}^{q}}}} 
                    {\left (\displaystyle \frac{1}{N} \displaystyle \sum_{e=1}^{N}\frac{1}{M^{2}}\sum_{i=1}^{M^{2}} \frac {f_{1}^{e}(n_{\rm{ie}})}{\epsilon_{i}}   \right )^q} ,
\label{eqeff2}
\end{equation}
and thus $F_{q}^{corr}$ is termed as efficiency corrected normalized moments. For the uniform (binomial) and the non-uniform (non-binomial) efficiencies in the acceptance region of the detectors, how efficiency corrections affect the NFM is studied. To create events with $80\%$ overall uniform efficiency to detect particles,  20 percent of the tracks are uniformly removed from the Toy model events. The events so obtained are termed here as the reconstructed-uniform (recU) events and have 80\% of the tracks that are there in the Toy model events. The analysis is performed on these events and using Eq.~\ref{def} NFM $F_{q}^{recU}$(M) are obtained.  

\begin{figure}
  \includegraphics[width=0.37\textwidth]{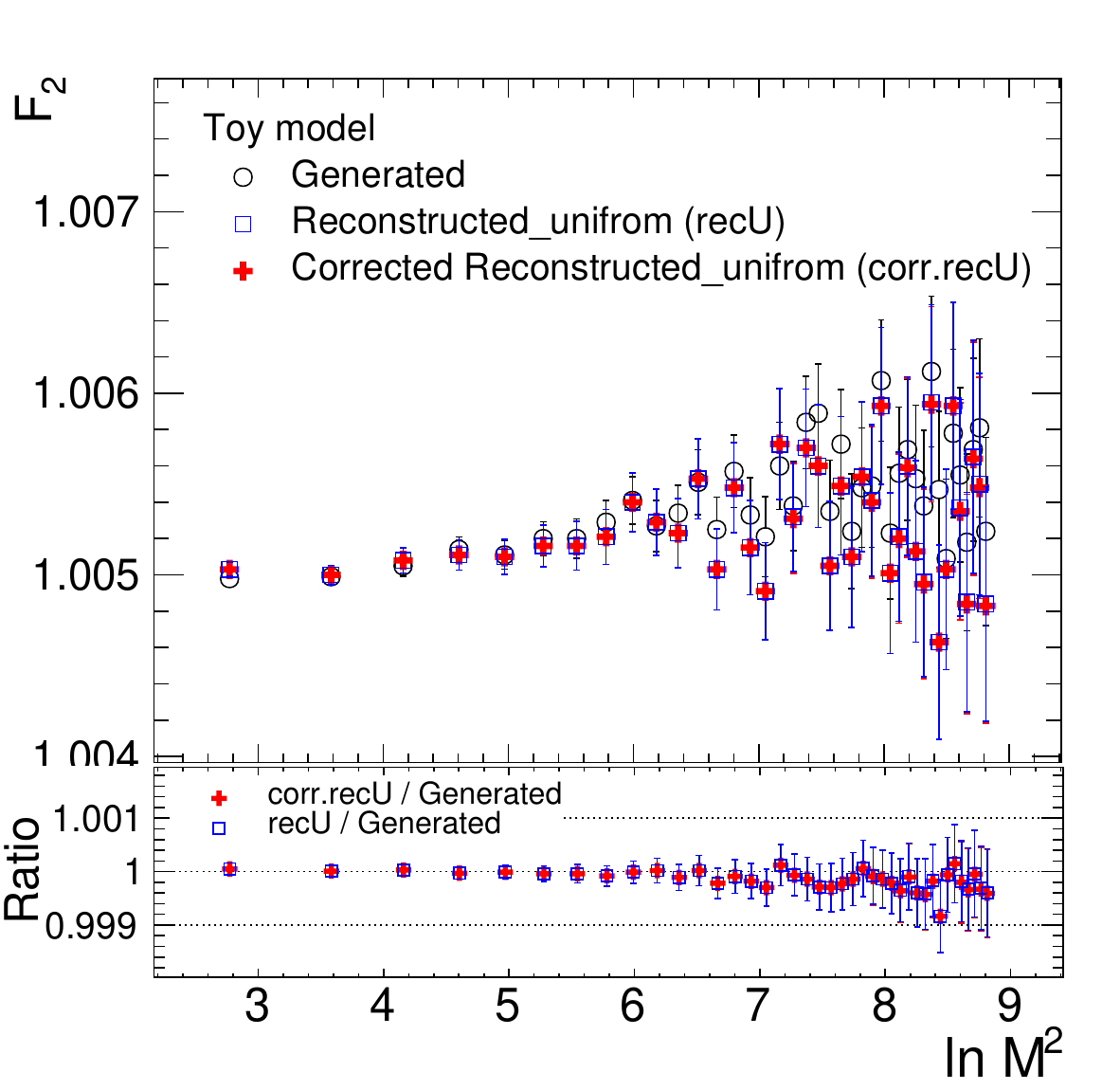}
   \caption{$F_{q=2}$ as function of $\ln M^{\rm{2}}$ for Toy model events, reconstructed-Uniform (recU) events and efficiency corrected  values of NFM for recU events. The lower panel shows the ratio graphs.}
  \label{effbin}
\end{figure}

Two-dimensional efficiency maps are calculated for all $M$ values such that for each $i^{{\rm th}}$ bin in the phase space that is divided in $M^{2}$ bins, there is an efficiency value $\epsilon_{i}$. For recU events corrected normalized factorial moments, $F_{q}^{corr.recU}$(M), are calculated using Eq.~\ref{eqeff2}. In Fig.~\ref{effbin}, M-scaling plots for q = 2 are given for these three cases. It is observed that  $F_{2}$(M) $\approx$  $F_{2}^{recU}$(M) $\approx$ $F_{\rm{2}}^{corr.recU}$(M). Therefore, the ratios $F_{\rm{2}}^{recU}$/$F_{2}$ and  $F_{\rm{2}}^{corr.recU}$/$F_{2}$  $\approx$ 1,  as shown in the lower panel of the figure. Thus, the true values of NFM are not affected when the efficiencies are of binomial in nature. In other words, normalized factorial moments are robust against binomial (uniform) efficiencies.

\begin{figure}
  \includegraphics[width=0.37\textwidth]{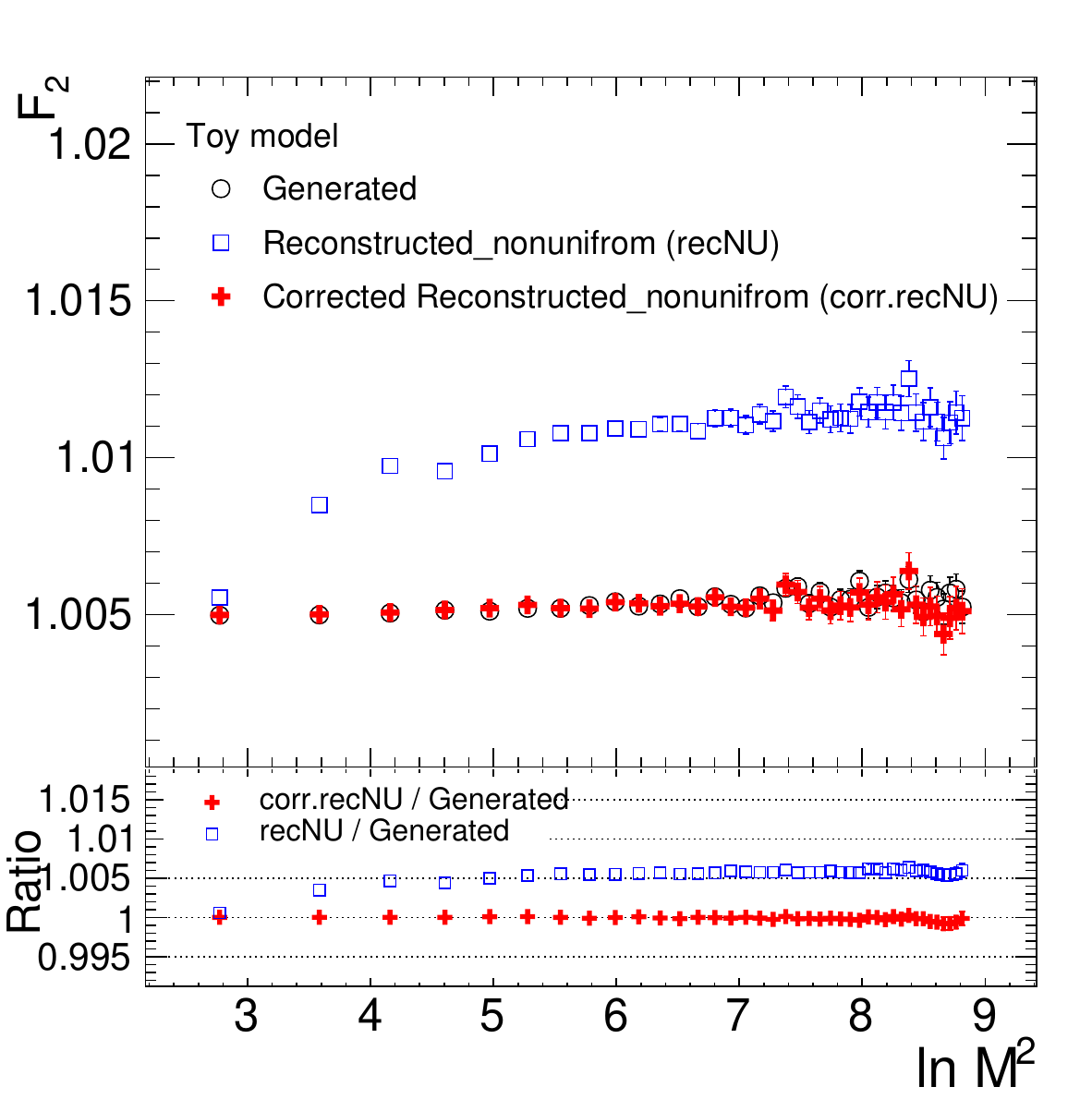}
  \caption{$F_{q=2}$ as function of $\ln M^{\rm{2}}$ for Toy model events and reconstructed-non Uniform (recNU) events.  Efficiency corrected $F_{q=2}$ values for recNU events are also shown. The lower panel shows the ratio plots.}
  \label{effnobin}
\end{figure}

For events with 80$\%$ non-uniform efficiencies, 20 percent of the tracks are removed from the selected phase space region of the Toy model events thereby resulting in a set of events with non-binomial efficiencies. The set of events so obtained are termed reconstructed-non uniform (recNU) events. The analysis is performed on recNU events and normalized factorial moments ($F_{q}^{recNU}$(M)) are calculated using Eq.~\ref{def}. As observed in Fig.~\ref{effnobin} it is observed that $F_{q}{\rm(M)}\!\not\equiv\!F_{q}^{recNU}$(M).

Efficiency maps are obtained for each $M$ value so as to have efficiency $\epsilon$ for each bin.  Corrected NFM ($F_{q}^{corr.recNU}$(M)) are calculated using Eq.~\ref{eqeff2}.   It is observed (Fig.~\ref{effnobin}) that $F_{q}{\rm(M)}\!\equiv\!F_{q}^{corr.recNU}$(M) and ratio $F_{q}^{corr.recNU}/F_{q} =1$ that  is shown with red markers in the lower panel. The true values of the normalized factorial moments are reproduced  by applying NFM formula with efficiency corrections. Thus,  in the case of non-binomial (non-uniform) detector efficiencies, Eq.~\ref{eqeff2} should be used to obtain the true values for NFM. 
 \section{Summary}
  \label{sum}
Normalized factorial moments (NFM) in the contours of \textit{intermittency} analysis are studied for the high multiplicity Toy model events simulated using  uniform distribution function. NFM do not show dependence on the number of bins. However, the methodology is found to be sensitive to the density fluctuations in the multiplicity distributions. A baseline value of the scaling exponent ($\nu$) from the Toy model events is  $\,1.603\pm0.016$, a value which is greater than the value predicted for the second-order phase transition in the Ginzburg-Landau formalism.  In addition, for the Toy model events $\nu$ is independent of the multiplicity but is dependent on the nature of distributions and hence the inherent fluctuations. $F_{q}$ is robust against the uniform efficiencies but for non-uniform efficiencies, calculations using efficiency-corrected NFM formula reproduce the true values of the observable. The observations and results obtained here serve as a baseline behavior of NFMs for future experimental investigations in this field.   
\section{Acknowledgement} 
R.G. is grateful to Rudolph C. Hwa and Edward Sarkisyan-Grinbaum for many fruitful discussions. The authors are thankful to Tapan K. Nayak, Igor Altsybeev, and Mesut Arslandok for the helpful discussions to complete this study. This work is partially funded by RUSA 2.0 grant sanctioned in favor of one of the authors by the Ministry of Education, Government of India.
\bibliographystyle{unsrt}
\bibliography{Toymodel1_RG}

\end{document}